%% file: main.tex
\documentclass{article}
\usepackage{spconf,amsmath,graphicx}

\input{./formatting/packages}
\input{./formatting/commands}

\input{./formatting/custom_formatting}
\input{./formatting/plots_and_diagrams}
\input{./formatting/custom_math_symbols}
\input{acronyms}

\title{Learned Decimation for Neural Belief Propagation Decoders\vspace{-10pt}}
\name{\begin{tabular}{c}
    (Invited Paper)\\[8pt]
    Andreas Buchberger\(^{\star}\),
    Christian H\"ager\(^{\star}\),
    Henry D. Pfister\(^{\dagger}\),\\
    Laurent Schmalen\(^{\ddagger}\), and
    Alexandre Graell i Amat\(^{\star}\)
  \end{tabular}\vspace{-12pt}\thanks{This work was partially funded by the Swedish Research Council under grant 2016-04253 and the Knut and Alice Wallenberg Foundation under grant 2018.0090.}}
\address{\normalsize \(^{\star}\) Dept. of Electrical Engineering, Chalmers University of Technology, Gothenburg, Sweden\\[-3pt]
        \normalsize \(^{\dagger}\) Dept. of Electrical and Computer Engineering, Duke University, Durham, North Carolina, USA\\[-3pt]
        \normalsize \(^{\ddagger}\) Communications Engineering Lab, Karlsruhe Institute of Technology (KIT), Karlsruhe, Germany  \vspace{-15pt}
}

\begin{document}

\maketitle
\begin{abstract}
We introduce a two-stage decimation process to improve the performance of \ac{NBP}, recently introduced by Nachmani \emph{et al.}, for short \ac{LDPC} codes.
In the first stage, we build a list by iterating between a conventional \ac{NBP} decoder and guessing the least reliable bit. The second stage iterates between a conventional \ac{NBP} decoder and learned decimation, where we use a neural network to decide the decimation value for each bit.  For a (128,64) \ac{LDPC} code, the proposed \ac{NBP} with decimation outperforms \ac{NBP} decoding by 0.75\,dB and performs within 1\,dB from \acl{ML} decoding at a block error rate of $10^{-4}$.
\end{abstract}
\acresetall
\vspace{-5pt}
\section{Introduction}
\Ac{BP} decoding can be formulated as a sparse deep \ac{NN} where instead of iterating between \acp{CN} and \acp{VN}, the messages are passed through unrolled iterations in a feed-forward fashion \cite{Nachmani2016,Nachmani2018}. For short block codes, the weights associated with the edges may counteract the effect of short cycles in the graph by scaling the messages accordingly. This is commonly referred to as \ac{NBP} and can be seen as a generalization of weighted \ac{BP} decoding where all individual messages are scaled by a single coefficient \cite{Chen2002}.
For a given code, the performance of \Ac{BP} and \ac{NBP} can be improved by using redundant parity-check matrices at the cost of increased complexity  \cite{Bossert1986,Kothiyal2005,Jiang2006,Halford2006,Hehn2010,Santi2018, Buchberger2020ISIT}. A pruning-based \ac{NBP} decoder was introduced in
\cite{Buchberger2020ISIT}, which improves the performance of \ac{NBP} at a lower decoding complexity.

It can be observed that the gain of \ac{NBP} over \ac{BP} for \ac{LDPC} codes is mostly in reducing the number of decoding iterations\textemdash the performance of \ac{NBP} and \ac{BP} converges for a sufficiently large number of iterations and  a fundamental gap between (N)BP and \ac{ML} decoding remains. Indeed, the error rate of (N)BP decoding of  \ac{LDPC} codes is dominated by absorbing sets from which the (N)BP decoder cannot recover. One way to improve performance is to run multiple decoders with different hard guesses for the least reliable bits\cite{Chase1972,PishroNik2004,Measson2008}. Each guess naturally splits the decoding process into two parallel decoders, one for each guess \cite{Measson2008}.
Similarly, in problems where multiple codewords can have similar posterior probabilities, such as lossy compression, one may make hard decisions for the most reliable bits without splitting the decoder, helping the \ac{BP} algorithm to converge to a codeword \cite{Filler2007,Wainwright2010,Castanheira2010,Aref2015}. We refer to both of these operations as \emph{decimation} though that term is typically associated with the second case.

In this work, we propose the use of decimation for \ac{NBP} decoding, thereby introducing a \ac{NBPD} decoder.
Our proposed decimation scheme consists of a \emph{list-based decimation} stage and a \emph{learned decimation} stage. We start with the list-based decimation stage and run a conventional \ac{NBP} decoder for \(\ell_\mathsf{max}\) iterations. We identify the least reliable \ac{VN}, i.e., the \ac{VN} with the lowest absolute \emph{a posteriori} \ac{LLR} and decimate it to \(\pm\infty\). Choosing the correct sign is essential as the correct sign will aid convergence  whereas the incorrect sign will hinder it. As the correct sign is unknown, we proceed with two graphs---one where the \ac{VN} is decimated to \(+\infty\) and one where the \ac{VN} is decimated to \(-\infty\). We iterate between decimating and decoding using the \ac{NBP} decoder for the desired number of times and end up with a list of codeword candidates. After the list-based decimation stage is complete, we continue with a learned decimation stage. Similar to the list-based decimation stage, we run a conventional \ac{NBP} decoder for \(\ell_\mathsf{max}\) iterations. For each \ac{VN}, we then use an \ac{NN} to decide to which value each \ac{VN} is decimated to. The sign of the \ac{VN} is not changed. We then iterate between the conventional \ac{NBP} decoder and the learned decimation for a desired number of decimation steps.
 We apply our proposed \ac{NBPD} decoder to an \ac{LDPC} code from the CCSDS standard and demonstrate a  performance within \(\SI{1}{\decibel}\) from \ac{ML} decoding.

 \section{Preliminaries}
 \label{sec:preliminaries}
 Consider a linear block code \(\mathcal{C}\) of length \(n\) and dimension \(k\) with parity-check matrix \(\bm{H}\) of size \(m \times n\), \(m \ge n - k\). The Tanner graph corresponding  to the parity-check matrix  \(\bm{H}\) is denoted as \(\mathcal{G} = (\mathcal{V}_\mathsf{v}, \mathcal{V}_\mathsf{c}, \mathcal{E})\), consisting of a set \mbox{\(\mathcal{V}_\mathsf{c}=\{\mathsf{c}_1,\ldots,\mathsf{c}_m\}\)} of \(m\) \acp{CN}, a set \(\mathcal{V}_\mathsf{v}=\{\mathsf{v}_1,\ldots,\mathsf{v}_n\}\) of \(n\) \acp{VN}, and a set of edges \(\mathcal{E}\) connecting \acp{CN} with \acp{VN}.
For each \ac{VN} \(\mathsf{v}\in \mathcal{V}_\mathsf{v}\) we define its neighborhood \mbox{\(\mathcal{N}(\mathsf{v}) \triangleq \left\{\mathsf{c}\in\mathcal{V}_\mathsf{c}:(\mathsf{v},\mathsf{c})\in \mathcal{E}\right\}\)}, i.e., the set of all \acp{CN} connected to \ac{VN} \(\mathsf{v}\), and equivalently, the neighborhood of a \ac{CN} \(\mathsf{c}\in \mathcal{V}_\mathsf{c}\) is defined as \(\mathcal{N}(\mathsf{c}) \triangleq \left\{\mathsf{v}\in\mathcal{V}_\mathsf{v}:(\mathsf{v},\mathsf{c})\in \mathcal{E}\right\}\).
Let \(\mu_{\mathsf{v}_i\rightarrow\mathsf{c}_j}^{(\ell)}\)  be the message passed from \ac{VN} \(\mathsf{v}_i\in\mathcal{V}_\mathsf{v}\) to \ac{CN} \(\mathsf{c}_j\in\mathcal{V}_\mathsf{c}\) and \(\mu_{\mathsf{c}_j\rightarrow\mathsf{v}_i}^{(\ell)}\) the message passed from \ac{CN} \(\mathsf{c}_j\in\mathcal{V}_\mathsf{c}\) to \ac{VN} \(\mathsf{v}_i\in\mathcal{V}_\mathsf{v}\) in the \(\ell\)-th decoding iteration. For \ac{BP} decoding, the \ac{VN} and \ac{CN} updates are
\begin{align}
  \mu_{\mathsf{v}_i\rightarrow\mathsf{c}_j}^{(\ell)} &=  \mu_{\mathsf{ch},\mathsf{v}_i} + \sum_{\mathsf{c}\in \mathcal{N}(\mathsf{v}_i)\backslash \mathsf{c}_j}  \mu_{\mathsf{c}\rightarrow\mathsf{v}_i}^{(\ell)}
  \label{eq:vn_update}
\end{align}
and
\begin{align}
  \mu_{\mathsf{c}_j\rightarrow\mathsf{v}_i}^{(\ell)} &= 2\tanh^{-1}\left(\prod_{\mathsf{v}\in \mathcal{N}(\mathsf{c}_j)\backslash \mathsf{v}_i}  \tanh\left(\frac{1}{2}\mu_{\mathsf{v}\rightarrow\mathsf{c}_j}^{(\ell)}\right)\right)
  \label{eq:cn_update}
\end{align}
respectively, where \(\mu_{\mathsf{ch},\mathsf{v}_i}\) is the channel message. For binary transmission over the additive white Gaussian noise channel
\begin{align*}
 \mu_{\mathsf{ch},\mathsf{v}_i} \triangleq \ln\left( \frac{p_{Y|B}(y_i|b_i=0)}{p_{Y|B}(y_i|b_i=1)}\right) \overset{}{=}  \frac{2y_i}{\sigma^2}
\end{align*}
where \(y_i\) is the channel output, \(b_i\) is the transmitted bit, and \(\sigma^2\) is the noise variance.
The \emph{a~posteriori} \ac{LLR} in the \(\ell\)-th iteration is
\begin{align*}
  \mu_{\mathsf{v}_i}^{(\ell)} &=  \mu_{\mathsf{ch},\mathsf{v}_i} + \sum_{\mathsf{c}\in \mathcal{N}(\mathsf{v}_i)}  \mu_{\mathsf{c}\rightarrow\mathsf{v}_i}^{(\ell)}.
\end{align*}

\subsection{Neural Belief Propagation}
For conventional \ac{BP}, the decoder iterates between \acp{VN} and \acp{CN} by passing messages along the connecting edges. For a given maximum number of iterations \(\ell_\mathsf{max}\),  one can \emph{unroll} the graph by stacking \(\ell_\mathsf{max}\) copies of the Tanner graph, and perform message passing over the unrolled graph. One way to counteract the effect of short cycles on the performance of \ac{BP} decoding for short linear block codes is to introduce weights for each edge of the unrolled Tanner graph \cite{Nachmani2016,Nachmani2018}. The resulting unrolled weighted graph can be interpreted as a \emph{sparse} \ac{NN} and accordingly decoding over the unrolled graph is referred to as  \ac{NBP}.  For \ac{NBP}, the update rules  \internalEq{eq:vn_update} and \internalEq{eq:cn_update} are modified to
\begin{align}
  \mu_{\mathsf{v}_i\rightarrow\mathsf{c}_j}^{(\ell)} &=w_{\mathsf{v}_i\rightarrow\mathsf{c}_j}^{(\ell)}\left( w_{\mathsf{ch},\mathsf{v}_i}^{(\ell)}\mu_{\mathsf{ch},\mathsf{v}_i} + \sum_{\mathsf{c}\in \mathcal{N}(\mathsf{v}_i)\backslash \mathsf{c}_j}  \mu_{\mathsf{c}\rightarrow\mathsf{v}_i}^{(\ell)}\right)
  \label{eq:vn_update_wbp}
\end{align}
and
\begin{align}
\mu_{\mathsf{c}_j\rightarrow\mathsf{v}_i}^{(\ell)} &= 2 w_{\mathsf{c}_j\rightarrow\mathsf{v}_i}^{(\ell)}\tanh^{-1}\left(\prod_{\mathsf{v}\in \mathcal{N}(\mathsf{c}_j)\backslash \mathsf{v}_i} \hspace{-12pt}\tanh\left(\frac{1}{2}\mu_{\mathsf{v}\rightarrow\mathsf{c}_j}^{(\ell)}\right)\hspace{-3pt}\right)
  \label{eq:cn_update_wbp}
\end{align}
where \( w_{\mathsf{ch},\mathsf{v}}^{(\ell)}\), \(w_{\mathsf{v}\rightarrow\mathsf{c}}^{(\ell)}\), and \(w_{\mathsf{c}\rightarrow\mathsf{v}}^{(\ell)}\), are the channel weights, the weights on the edges connecting \acp{VN} to \acp{CN}, and the weights on the edges connecting \acp{CN} to \acp{VN}, respectively.
The \emph{a~posteriori} \ac{LLR} in the \(\ell\)-th iteration is
\begin{align*}
  \mu_{\mathsf{v}_i}^{(\ell)} &= w_{\mathsf{ch},\mathsf{v}_i}^{(\ell)}\mu_{\mathsf{ch},\mathsf{v}_i} + \sum_{\mathsf{c}\in \mathcal{N}(\mathsf{v}_i)}  \mu_{\mathsf{c}\rightarrow\mathsf{v}_i}^{(\ell)}.
\end{align*}
In \internalEq{eq:vn_update_wbp} and \internalEq{eq:cn_update_wbp} the weights are untied over all nodes as well as over all iterations, i.e., each edge has an individual weight. In order to reduce complexity and storage requirements for \ac{NBP}, the weights can also be tied. In \cite{Nachmani2018,Lian2019}, tying the weights temporally, i.e., over  iterations, and spatially, i.e., all edges within a layer have the same weight, was explored.

For finding the edge weights, one may consider a binary classification task for each of the \(n\) bits. As a loss function, the average bitwise cross-entropy between the transmitted bits and the  output \acp{LLR} of the final \ac{VN} layer can be used \cite{Nachmani2016,Nachmani2018}.
The optimization behavior can be improved by using a multiloss function \cite{Nachmani2016,Nachmani2018}, where the overall loss is the average  bitwise cross-entropy between the transmitted bits and the  output \acp{LLR} of each \ac{VN} layer. Using the loss function, stochastic gradient descent (and variants thereof) can be used to optimize the weights.

\section{Decimated Neural Belief Propagation Decoder}
While \ac{NBP} improves upon \ac{BP} decoding for a fixed (relatively small) number of iterations, for \ac{LDPC} codes \ac{NBP} and  \ac{BP} appear to yield the same performance for large enough  number of iterations, and a gap to the \ac{ML} performance remains. To overcome this limitation of \ac{NBP}, here we propose a two-stage decimation process on top of \ac{NBP} decoding consisting of a \emph{list-based decimation} stage and a \emph{learned decimation} stage. In both stages, we iterate between a decimation process and a conventional \ac{NBP} decoder with \(\ell_\mathsf{max}\) iterations for which we consider the case where the weights are tied over  iterations, i.e, \(w_{\mathsf{ch},\mathsf{v}_i}^{(\ell)} = w_{\mathsf{ch},\mathsf{v}_i}\), \(w_{\mathsf{v}_i\rightarrow\mathsf{c}_j}^{(\ell)} = w_{\mathsf{v}_i\rightarrow\mathsf{c}_j}\), and \(w_{\mathsf{c}_j\rightarrow\mathsf{v}_i}^{(\ell)} = w_{\mathsf{c}_j\rightarrow\mathsf{v}_i}\), and are  optimized  for a large number of iterations.
We denote our proposed decoder as \(\text{NBP-D}(\ell_\mathsf{max}, n_\mathsf{D}, n_\mathsf{LD})\), where \(\ell_\mathsf{max}\) denotes the number of iterations of the conventional \ac{NBP} decoder, \(n_\mathsf{D}\) is the number of decimations in the list-based decimation stage, and \(n_\mathsf{LD}\) denotes the number of decimations in the learned decimation stage. In the following, we describe the two decimation stages in detail, provide the training procedure, and give a brief complexity discussion.

\subsection{List-based Decimation Stage}
\label{sec:list_decimation}
\begin{figure}
  \centering
  \includegraphics{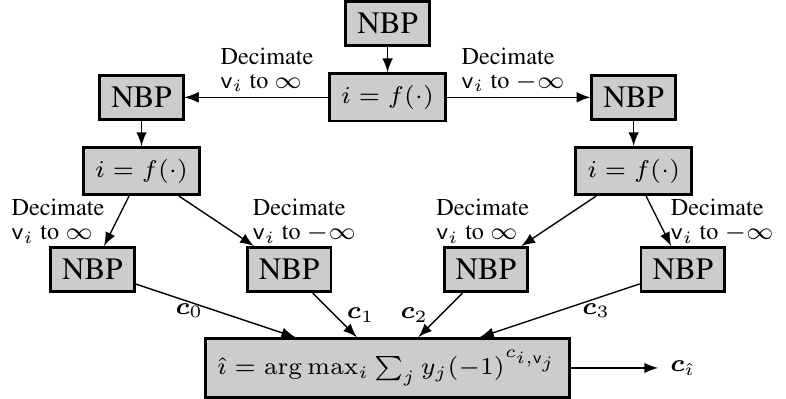}
  \vspace{-5pt}
  \caption{Decoding tree for \(n_\mathsf{D} = 2\) and \(n_\mathsf{LD} = 0\) where \(f(\cdot) = \arg\min_j{|\mu^{(\ell_\mathsf{max})}_{\mathsf{v}, j}}|\).\vspace{-15pt}}
  \label{fig:decoding_tree}
\end{figure}

We start by running \ac{NBP} for \(\ell_\mathsf{max}\) iterations. We  then identify the \ac{VN} \(\mathsf{v}\) with the lowest absolute  \emph{a~posteriori} \ac{LLR} \(\mu_\mathsf{v}^{(\ell_\mathsf{max})}\), i.e., the least reliable  \ac{LLR} and decimate it to \(\pm\infty\), i.e., set \(\mu_{\mathsf{ch},\mathsf{v}} = \pm\infty\). Since the correct sign is unknown, we build a decoding tree: Each time we decimate a \ac{VN}, we create two new graphs where in one the \ac{VN} is decimated to \(+\infty\) and in the other the \ac{VN} is decimated to \(-\infty\). We then run the \ac{NBP} decoder for each of the two graphs. We alternate between running \ac{NBP} and decimation \(n_\mathsf{D}\)  times. In \internalFig{fig:decoding_tree}, we depict the decoding tree for \(n_\mathsf{D} = 2\).

\subsection{Learned Decimation Stage}
\label{sec:learned_decimation}
While using the previously described list-based decimation allows to significantly boost the performance, it has the drawback that the complexity increases exponentially in the number of decimations. To reduce the size of the decoding tree, we propose an \ac{NN}-based approach. More precisely, instead of continuing to unfold the decoding tree, we use an \ac{NN} to decide  to which value  each \ac{VN}  should be decimated. The \ac{NN} takes the incoming messages to the node and the channel \ac{LLR} as  input and outputs a value whose absolute value  is then added to the channel message. The sign is kept according to the output \ac{LLR} of the respective node. Hence, we decimate each \ac{VN} \(\mathsf{v}\in\mathcal{V}_\mathsf{v}\) according to
\begin{align*}
  \mu_{\mathsf{ch}, \mathsf{v}} &= \mu_{\mathsf{ch}, \mathsf{v}} + \sign{\mu_{\mathsf{v}}^{(\ell_\mathsf{max})}}\nonumber\\ &\qquad\qquad\cdot\left|f_{\mathsf{NN}}\left(\mu_{\mathsf{ch}, \mathsf{v}}, \{\mu_{\mathsf{c}\rightarrow\mathsf{v}}^{(\ell_\mathsf{max})}| \mathsf{c}\in\mathcal{N}(\mathsf{v})\}, \bm{\theta}\right)\right|,
\end{align*}
where \(f_\mathsf{NN}(\cdot)\) denotes the \ac{NN} and \(\bm{\theta}\) denotes all trainable parameters of the \ac{NN}.
The same \ac{NN} is applied for all \acp{VN} and weights are shared between \acp{VN} and all decimation steps. We alternate learned decimation and running \ac{NBP} \(n_\mathsf{LD}\) times.

After the \(n_\mathsf{D}\) list-based decimations and \(n_\mathsf{LD}\) learned decimations, we obtain \(2^{n_\mathsf{D}}\) codeword candidates \(\bm{c}_i\) and we choose the most likely codeword.
The \ac{NBPD} decoder is described in \internalAlgorithm{lst:ldnbp}.

\begin{algorithm}[t]
 \caption{\Acl{NBPD}.}
 \label{lst:ldnbp}
 \small
 \begin{algorithmic}[1]
   \algrenewcommand\algorithmicindent{0.75em}%
   \renewcommand\algorithmicdo{\textbf{:}}
   \renewcommand\algorithmicrequire{\textbf{Input:}}
   \renewcommand\algorithmicthen{\textbf{:}}
   \algnewcommand{\LineComment}[1]{\Statex {\vspace{-7pt}\hrulefill\vspace{-1pt}}\Statex{ \(\triangleright\) \textit{#1}}}
   \algnewcommand{\AlgoLine}{\Statex {\vspace{-8pt}\hrulefill\vspace{-1pt}}}
   \algnewcommand{\StatexIndent}[1]{\Statex{\hspace{\algorithmicindent}\hspace{\algorithmicindent}\hspace{\algorithmicindent}\hspace{\algorithmicindent}#1}}

\Require Trained NBP decoder  with \(\ell_\mathsf{max}\) iterations
  \Statex{ Channel outputs \(y_i\) and \(\mu_{\mathsf{ch},\mathsf{v}_i}\) for all \acp{VN} \(\mathsf{v}_i\in\mathcal{V}_\mathsf{v}\)}
  \Statex{ \# of decimations in the list-based decimation stage \(n_\mathsf{D}\)}
  \Statex{ \# of decimations in the learned decimation stage \(n_\mathsf{LD}\)}

\AlgoLine
   \State{\(\mathcal{M}_1 \leftarrow \{\mu_{\mathsf{ch},\mathsf{v}}\}\) and \(\mathcal{M}_i\leftarrow\{\}\,\forall\,i\in\{2,\ldots, 2^{n_\mathsf{D}}\}\)}

  \LineComment{\nameref{sec:list_decimation} (\internalLinkSection{sec:list_decimation})}
  \For {\(i\leftarrow\{1, \ldots, n_\mathsf{D}\}\)}
    \For{\(j\leftarrow\left\{1,\ldots,2^{i-1}\right\}\)}
      \State {Decode \(\mathcal{M}_j\) using NBP to \(\mu_\mathsf{v}^{(\ell_\mathsf{max})}\)}
      \State {\(k\leftarrow\arg\min_i | \mu_{\mathsf{v}_i}^{(\ell_\mathsf{max})}|\)}
      \State {\(\mathcal{M}_{2^{i-1}+j} \leftarrow \mathcal{M}_j\)}
      \State {\({\mu}_{\mathsf{ch},\mathsf{v}_k} \leftarrow +\infty\) with \({\mu}_{\mathsf{ch},\mathsf{v}_k} \in \mathcal{M}_j\)}
      \State {\({\mu}_{\mathsf{ch},\mathsf{v}_k} \leftarrow -\infty\) with \({\mu}_{\mathsf{ch},\mathsf{v}_k} \in \mathcal{M}_{2^{i-1}+j}\)}
    \EndFor
  \EndFor
  \LineComment{\nameref{sec:learned_decimation} (\internalLinkSection{sec:learned_decimation})}
  \For{\(i\leftarrow\{1, \ldots, n_\mathsf{LD}\}\)}
    \For{\(j\leftarrow\{1, \ldots, 2^{n_\mathsf{D}}\}\)}
      \State {Decode \(\mathcal{M}_j\) using NBP  to \(\mu_\mathsf{v}^{(\ell_\mathsf{max})}\)}
        \For{\(\mathsf{v}\in\mathcal{V}_\mathsf{v}\)}
          \State{\(\mu_{\mathsf{ch}, \mathsf{v}} \leftarrow \mu_{\mathsf{ch}, \mathsf{v}}\)}
          \Statex{\hspace{3.50em}\( + \sign{\mu_{\mathsf{v}}} \left|f_{\mathsf{NN}}\left(\mu_{\mathsf{ch}, \mathsf{v}}, \{\mu_{\mathsf{c}\rightarrow\mathsf{v}}^{(\ell_\mathsf{max})}, \mathsf{c}\in\mathcal{N}(\mathsf{v})\}, \bm{\theta}\right)\right|\)}
      \EndFor

    \EndFor
  \EndFor
  \AlgoLine
  \For{\(j\leftarrow\{1, \ldots, 2^{n_\mathsf{D}}\}\)}
  \State {Decode \(\mathcal{M}_j\) using NBP  to \(\mu_\mathsf{v}^{(\ell_\mathsf{max})}\)}
\State{\mbox{\(\bm{c}_j \leftarrow \left(c_{j, \mathsf{v}_1}, \ldots, c_{j, \mathsf{v}_n}\right)\), \(c_{j, \mathsf{v}}=\left(1-\sign{\mu_{\mathsf{v}}^{(\ell_\mathsf{max})}}\right)/2\)}}
  \EndFor
  \State {\(\hat{\imath} = \arg \max_i \sum_j y_j(-1)^{c_{i,\mathsf{v}_j}}\)}
  \Statex {\textbf{return} \(\bm{c}_{\hat{\imath}}\)}
 \end{algorithmic}
 \end{algorithm}

\subsection{Training}
We perform training in two stages. We first consider a conventional \ac{NBP} decoder with weights tied over the iterations. We train this decoder using the multiloss bitwise cross-entropy between the output \acp{LLR} of each \ac{VN} layer and the correct codeword as the loss function and the Adam optimizer \cite{Nachmani2018}. We then freeze the learned weights and unroll the \ac{NBPD}(\(\ell_\mathsf{max}, n_\mathsf{D}, n_\mathsf{LD}\)) decoder. To reduce the complexity of the training procedure, we only consider the correct path through the decoding tree in the list-based decimation stage, i.e., we assume a genie that provides the correct sign of the decimated \ac{VN}. This incurs no loss in \ac{BLER} as, of the $2^{n_\mathsf{D}}$ decoders, only the one with the correct decisions can return the correct codeword. We train the unrolled \ac{NBPD} decoder using the multiloss bitwise cross-entropy between the output \acp{LLR} of each \ac{VN} layer and the correct codeword as the loss function and the Adam optimizer.

\subsection{Complexity Discussion}
On a high level, by disregarding hardware implementation details, the \ac{CN} update is the most complex operation due to the evaluation  of the  \(\tanh\) and inverse \(\tanh\) functions. A commonly used complexity measure is given by \(\overline{d}_\mathsf{c} m\ell_\mathsf{max}\)\cite{Smith2010}, where \(\overline{d}_\mathsf{c}\) is the average \ac{CN} degree and \(m\) the number of \acp{CN}. The complexity of the \ac{NBPD} decoder follows as
\begin{align*}
   \overline{d}_\mathsf{c}m\ell_\mathsf{max}\left( 2^{n_\mathsf{D}+1} - 1  + n_\mathsf{LD}2^{n_\mathsf{D}}\right).
\end{align*}

The memory requirement for the decoders is dominated by the number of weights that need to be stored. For the \ac{NBP} decoder, this corresponds to the number of edges in the Tanner graph. In the case of \ac{NBPD}, additionally the weights of the \ac{NN} need to be taken into account. Note that the weights also entail additional multiplications.

\begin{figure}
  \centering
  \includegraphics{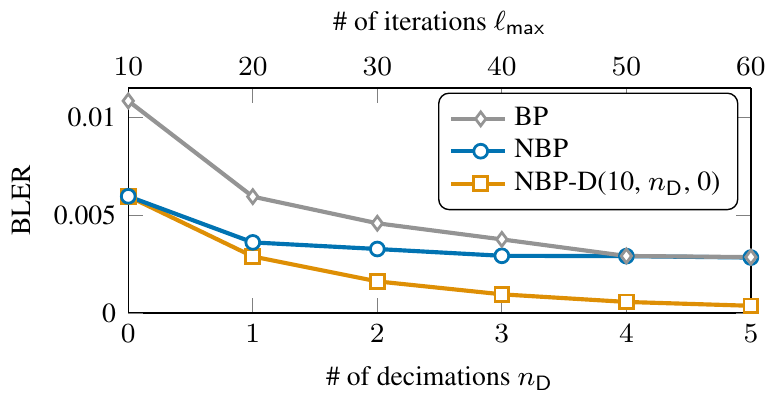}
  \vspace{-14pt}
  \caption{BLER at \(E_\mathsf{b}/N_0 = \SI{4}{\decibel}\).}
  \label{fig:bler_vs_list_size}
    \vspace{-10pt}
\end{figure}
\section{Numerical Results}
\vspace{-5pt}
In this work, we consider an \ac{LDPC} code of length \(n=128\) and rate \(0.5\) with average \ac{CN} degree \(\bar{d}_\mathsf{c} = 8\)  as defined by the CCSDS standard. Its parity-check matrix is of size \(64\times 128\). It is important to note that the presented concepts are not limited to a specific code and extend to any other sparse code.

For training the \ac{NBP} and  \ac{NBPD} decoders, the batch size was set to \(128\), the learning rate to \(0.001\), and the Adam optimizer was used for gradient updates. The \ac{NN} of the \ac{NBPD} decoder consists of three layers where the first and second layer contain \(16\) neurons and use the ReLU activation function, and the third layer consists of a single neuron with no activation function (i.e., a linear layer).

In \internalFig{fig:bler_vs_list_size}, we plot the \ac{BLER} as a function of the number of iterations for conventional \ac{BP},  \ac{NBP}, and \ac{NBPD} for which \(n_\mathsf{LD} = 0\), i.e., no learned decimation is employed. For a fixed number of iterations, \ac{NBP} outperforms \ac{BP}. For \ac{LDPC} codes, the gain of \ac{NBP} over \ac{BP} appears to vanish with an increasing number of iterations, and at \(50\) iterations \ac{BP} and \ac{NBP} have virtually the same performance. Importantly, increasing the number of iterations even further does not result in an improved performance. Considering the \ac{NBPD} decoder, the decimation enables us to outperform (N)BP, even for a single decimation. It is important to note that while the complexity of (N)BP increases linearly in the number of iterations, the complexity of \ac{NBPD} increases exponentially in the number of decimations \(n_\mathsf{D}\).
\begin{figure}[t]
  \centering
  \includegraphics{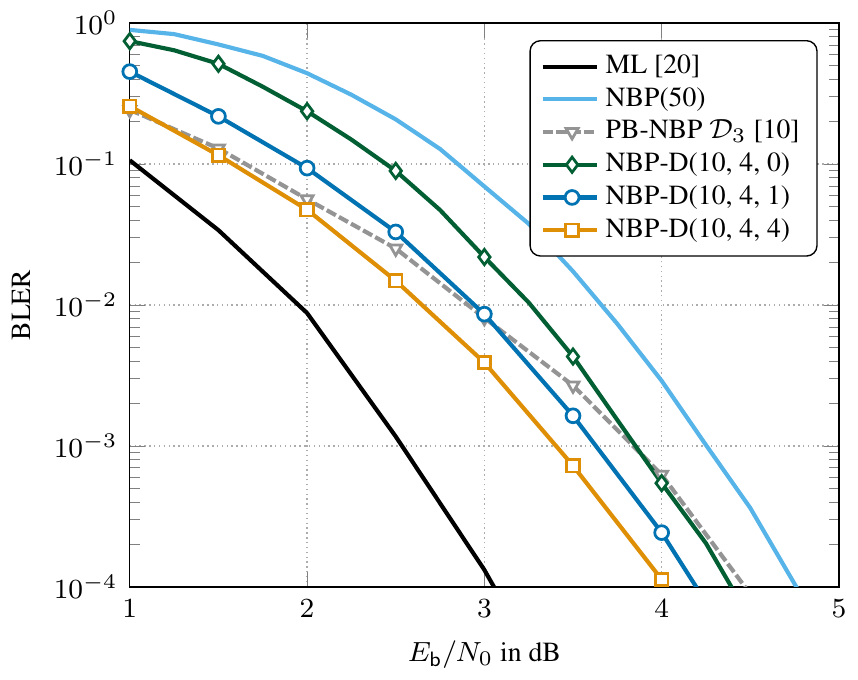}
  \vspace{-25pt}
  \caption{BLER for the CCSDS \((128, 64)\) \ac{LDPC} code.\nocite{channelcodes}}
  \label{fig:bler_ccsds}
\end{figure}

In \internalFig{fig:bler_ccsds}, we show the \ac{BLER} as a function of \(E_\mathsf{b}/N_0\). \Ac{NBP} with \(50\) iterations (\(\text{NBP}(50)\)) performs about \(\SI{1.75}{\decibel}\) from \ac{ML}. Increasing the number of iterations even further would not result in a better performance (see  \internalFig{fig:bler_vs_list_size}).  \(\text{NBP-D}(10, 4, 0)\)  improves by \(\SI{0.4}{\decibel}\) over \(\text{NBP}(50)\) and matches the performance of pruning-based NBP \(\mathcal{D}_3\) in \cite{Buchberger2020ISIT}. Adding a single learned decimation (\(\text{NBP-D}(10, 4, 1)\)) improves the performance by \(\SI{0.3}{\decibel}\). Allowing four learned decimation steps (\(\text{NBP-D}(10, 4, 4)\)), the gap to \ac{ML} is further reduced to  slightly less than \(\SI{1}{\decibel}\). The choice of \(n_\mathsf{D}=4\) is a trade-off between performance and complexity. Increasing \(n_\mathsf{D}\rightarrow n\) will eventually lead to \ac{ML} decoding. Relying solely on learned decimation (\(n_\mathsf{D}=0\)) is not competitive as the correct choice of the sign for \acp{VN} of low reliability is never guaranteed.
In \internalTab{tab:complexity}, we compare the complexities for the decoders in \internalFig{fig:bler_ccsds}. The numbers in parentheses indicate how much more complex a decoder is with reference to \(\text{NBP}(50)\).
\begin{table}[t]
  \centering
  \vspace{-10pt}
  \caption[Complexity]{Complexity of the decoders in \internalFig{fig:bler_ccsds}. The value in parentheses indicates how many times more complex the decoder is compared to \ac{NBP} with \(50\) iterations.}
  \label{tab:complexity}
  \begin{tabularx}{\columnwidth}{Xrlr}
    \toprule
     &\multicolumn{2}{c}{Complexity} & \# of weights\\
    \midrule
    \ac{NBP}(\(50\)) & \(25600\) & (\(1.0\)) & \(1152\)\\
    PB-NBP \(\mathcal{D}_3\) \cite{Buchberger2020ISIT} & \(25920\) & (\(1.0\)) & \(28416\)\\
    \midrule
    \ac{NBPD}(\(10,4, 0\)) & \(138240\) & (\(5.4\)) & \(1152\)\\
    \ac{NBPD}(\(10,4,1\)) & \(220160\) & (\(8.6\)) & \(1553\)\\
    \ac{NBPD}(\(10,4,4\)) & \(465920\)  & (\(18.2\)) & \(1553\)\\
    \bottomrule
  \end{tabularx}
  \vspace{-8pt}
\end{table}

\section{Conclusion}
\acreset{NBP,NBPD}
\vspace{-5pt}
We  proposed a \ac{NBPD} decoder for \ac{LDPC} codes. By combining a list-based decimation stage and a learned decimation stage where a neural network  learns which \acp{VN} to decimate, we have shown that we can significantly improve the performance of \ac{BP} and \ac{NBP}, and achieve a performance within \(\SI{1}{\decibel}\) from \ac{ML}. The concept of \ac{NBPD} can be applied to any other short block code. For sparse codes, similar performance improvements can be expected.

\bibliographystyle{IEEEbib}
\bibliography{IEEEabrv,CONFabrv,literature}
\end{document}

%% file: formatting/packages.tex
\usepackage[final]{listings}
\usepackage{acro}

\usepackage{xifthen}

\usepackage{cite}

\usepackage{nohyperref}
\usepackage{nameref}
\usepackage{tikz}
\usepackage{pgfplots}
\usetikzlibrary{arrows, positioning, shapes, calc, spy}
\usepgfplotslibrary{fillbetween, groupplots}

\usepackage{amsmath}
\usepackage{amssymb,amsfonts,amsthm}

\usepackage{bm}

\usepackage{siunitx}

\usepackage{ctable}

\usepackage{xcolor}

\usepackage{algorithm,algpseudocode}

\usepackage{ifthen}

\usepackage{marginnote}

%% file: formatting/commands.tex
\newcommand{\internalLinkSection}[1]{Section \ref{#1}}

\newcommand{\internalTab}[1]{Table\,{\ref{#1}}}
\newcommand{\internalFig}[1]{Fig.\,{\ref{#1}}}
\newcommand{\internalEq}[1]{({\ref{#1}})}

\newcommand{\internalAlgorithm}[1]{Algorithm\,{\ref{#1}}}

%% file: formatting/custom_formatting.tex
\let\originalleft\left
\let\originalright\right
\renewcommand{\left}{\mathopen{}\mathclose\bgroup\originalleft}
\renewcommand{\right}{\aftergroup\egroup\originalright}

%% file: formatting/plots_and_diagrams.tex
\definecolor{curve_color00}{HTML}{000000}
\definecolor{curve_color01}{HTML}{0173B2}
\definecolor{curve_color02}{HTML}{DE8F05}
\definecolor{curve_color03}{HTML}{ECE133}
\definecolor{curve_color04}{HTML}{025E33}
\definecolor{curve_color05}{HTML}{56B4E9}
\definecolor{curve_color06}{HTML}{FBAFE4}
\definecolor{curve_color08}{HTML}{949494}
\definecolor{curve_color10}{HTML}{56B4E9}

\definecolor{block_gray}{RGB}{204,204,204}
\definecolor{block_background_gray}{RGB}{240,240,240}
\definecolor{block_background_line_gray}{RGB}{175,175,175}
\definecolor{block_weak_gray}{RGB}{240,240,240}

\tikzstyle{line}=[draw, thick]
\tikzstyle{arrow}=[draw, -latex, thick, midway]
\tikzstyle{line}=[draw, -, thick, midway]

\tikzstyle{block} = [rectangle, draw, thick, fill=block_gray,
    text width=2.75em, text centered, rounded corners, minimum height=1.25em]
\tikzstyle{block_circle} = [circle, draw, thick, fill=block_gray, text centered]
\tikzstyle{channel} = [cloud, draw,thick, cloud puffs=12, cloud puff arc=120, aspect=2, fill=block_gray]

\tikzstyle{block_background} = [rectangle, draw, align=center, thick, draw=block_background_line_gray, fill=block_background_gray, rounded corners]

\pgfplotsset{
  legend style={font=\footnotesize},
  every tick label/.append style={font=\footnotesize},
  every axis label/.append style ={font=\footnotesize},
	legend image code/.code={
		\draw[mark repeat=2,mark phase=2]
			plot coordinates {
				(0cm,0cm)
				(0.3cm,0cm)        
				(0.55cm,0cm)         
			};%
	}
}

%% file: formatting/custom_math_symbols.tex
\newcommand{\sign}[1]{\operatorname{sign}\left(#1\right)}

\DeclareSIUnit{\belmilliwatt}{Bm}
\DeclareSIUnit{\dBm}{\deci\belmilliwatt}
\DeclareSIUnit{\bit}{bit}
\DeclareSIUnit{\bits}{bits}

%% file: acronyms.tex
\DeclareAcronym{BCH}{
	short = BCH,
	long= Bose-Chaudhuri-Hocquenghem
}

\DeclareAcronym{BER}{
	short = BER,
	long= bit error rate
}

\DeclareAcronym{BLER}{
	short = BLER,
	long= block error rate
}

\DeclareAcronym{BP}{
	short = BP,
	long = belief propagation
}

\DeclareAcronym{CN}{
	short = CN,
	long = check node
}

\DeclareAcronym{NBPD}{
	short = \mbox{NBP-D},
	long = neural belief propagation with decimation
}

\DeclareAcronym{LDPC}{
	short = LDPC,
	long = low-density parity-check
}

\DeclareAcronym{LLR}{
	short = LLR,
	long = log-likelihood ratio
}

\DeclareAcronym{MBBP}{
	short = MBBP,
	long = multiple-bases belief propagation
}

\DeclareAcronym{NBP}{
	short = NBP,
	long = neural belief propagation
}

\DeclareAcronym{NN}{
	short = NN,
	long = neural network
}

\DeclareAcronym{NOMS}{
	short = NOMS,
	long = neural offset min-sum
}

\DeclareAcronym{ML}{
	short = ML,
	long = maximum-likelihood
}

\DeclareAcronym{relu}{
	short = reLu,
	long = rectified linear unit
}

\DeclareAcronym{RM}{
	short = RM,
	long= Reed-Muller
}

\DeclareAcronym{RNN}{
	short = RNN,
	long= recurrent neural network
}

\DeclareAcronym{STE}{
	short = STE,
	long = straight-through estimator
}

\DeclareAcronym{PBNBP}{
	short = PB-NBP,
	long = pruning-based neural belief propagation
}

\DeclareAcronym{PBNOMS}{
	short = PB-NOMS,
	long = pruning-based neural offset min-sum
}

\DeclareAcronym{VN}{
	short = VN,
	long = variable node
}